\documentclass[showpacs,prd,twocolumn,aps,amsmath,amssymb,unsortedaddress]{revtex4}
\usepackage[dvips]{graphicx}
\usepackage{dcolumn}
\usepackage{amsmath}
\usepackage{times}

\begin{document}

\title{Algebraic Solution of the  Harmonic Oscillator With Minimal
Length Uncertainty Relations}
\author{K.\ Gemba, Z.\ T.\ Hlousek and Z.\ Papp}
\affiliation{ Department of Physics and Astronomy, California State
University, Long Beach, California 90840 }

\date{\today}

\begin{abstract}\noindent
In quantum mechanics with minimal length uncertainty relations the Heisenberg-Weyl algebra of the one-dimensional harmonic oscillator is a deformed $SU(1,1)$ algebra. The eigenvalues and eigenstates are constructed algebraically and they form the infinite-dimensional representation of the deformed $SU(1,1)$ algebra. Our construction is independent of prior knowledge of the exact solution of the Schr\"odinger equation of the model.
The approach can be generalized to the $D$-dimensional oscillator with non-commuting coordinates.

\end{abstract}

\maketitle

\section{\label{intro}Introduction}

Uncertainty relations are one of the pillars of quantum physics. They are directly
related to the basic commutator relations and to quantum equations of motion.
In ordinary quantum mechanics the basic commutator between the position and momentum
operators in one dimension is given by (we use units such that $\hbar = 1$),
\begin{equation}\label{eq:zero}
\left[ x,p\right] = i~.
\end{equation}
In this paper we shall consider the quantum mechanics where Eq. (\ref{eq:zero}) is modified or, in modern language, deformed.
Modified uncertainty relations appear in many different  areas of physics, sometimes directly and sometimes
in disguise. 
For example, in a system such as a complex molecule, there are length scales below which the
physics is complicated and some effective description is sufficient.
It is possible to capture some of the effective physics by a modification of the uncertainty relations. 
Rotational and vibrational states of molecules and deformed nuclei can be described
using models with deformed basic commutators. Similar applications also appear in the physics of deformed heavy nuclei.
Another area where modified basic commutators play some role is quantum optics. Various entangled and squeezed
coherent states are modeled successfully using this approach. Even more, such states can be experimentally
realized. Entangled states are also of importance in quantum computing.
From our perspective it is most exciting that
quantum theory of gravity requires that basic commutators of the quantum mechanics
be altered. This seems to be the case in both, the loop quantum gravity and in the string theory.
Very energetic test particles for probing very small scales on the order of the Planck
length disturb gravitationally the very space-time they are probing. The effect is captured as a
modification of the position-momentum uncertainty relation. Modified uncertainty relations
require modified basic commutators and imply the existence of minimal
length and minimal momentum.  There are also examples of
modified special relativity theory with invariant minimal length or minimal momentum, or both.

Uncertainty relations have a profound consequence on physics. For example,
position-momentum uncertainty relations in the ordinary quantum mechanics,
\begin{equation}
\Delta x\Delta p \geq \frac{1}{2}
\end{equation}
 reflect directly the basic commutator relation, Eq. (\ref{eq:zero}).
In the ordinary quantum mechanics it is possible
to construct states with zero uncertainty in position or momentum (of course, not simultaneously).
In other words, the space-time is a sharp continuum. Within the framework of the
ordinary quantum mechanics the usual uncertainty relations imply that it should be possible to measure,
at least in principle, the position and the momentum with absolute certainty, of course not at the same time!
On the other hand if the theory is endowed, for example,
with minimal length by modifying uncertainty relations, then the position is no longer a
viable observable and is called fuzzy.  We loose the Schr\"odinger equation as a differential or integral
equation in spatial coordinates. Coordinate representation of ordinary quantum mechanics becomes some kind of
effective description valid at sufficiently large scale.
There is a profound effect on the spectrum of states and on the scattering properties of systems in the modified
quantum theory.
If the theory is endowed with both, minimal length and
minimal momentum we loose both, coordinate and the momentum representations, so we are left with representation-free operator methods.  Both spectral and scattering
properties of systems deviate greatly from that described by the ordinary quantum mechanics.
Perhaps some experiments can be devised to look for and to measure such discrepancies an to
determine the size of deformation parameters.

The purpose of this paper is to study the energy eigenvalues and eigenvectors
of the one-dimensional and $D$-dimensional isotropic harmonic
oscillator model in the quantum mechanics with minimal length uncertainty relations.
In Ref.\ \cite{kempf+mm}, the energy eigenvalues of the one-dimensional 
harmonic oscillator with minimal length uncertainty relations
were calculated by solving the Schr\"odinger equation in momentum space.
In Ref.\ \cite{minic-atall} it was shown, again by solving the Schr\"odinger equation
in momentum space, that the wave-functions are given by Gegenabuer polynomials \cite{gradshteyn}.
In Ref.\ \cite{meljanac} ladder operators for the model were constructed by
using the knowledge of the exact wave functions and energy eigenvalues and the
recursion relations of the Gegenbauer polynomials.

In this paper we present a complete solution of the one-dimensional harmonic
oscillator in quantum theory with minimal length uncertainties.
We make no use of the knowledge of exact energy eigenvalues and wave-functions.
We show that the Heisenberg-Weyl algebra of the model is a
deformed $SU(1,1)$ algebra. We arrive at this algebra by showing that the model
is equivalent to a symmetric P\"oschl-Teller model. Operators that realize this deformed $SU(1,1)$ serve as
the ladder operators for the harmonic oscillator model in the minimal length quantum theory.
We can repeat the construction for the isotropic oscillator in $D$-dimensions.
It is worth mentioning that the $D$-dimensional
quantum mechanics with minimal length also features non-commuting coordinates.

The paper is organized as follows. In Section {\ref{sec2}}, following
\cite{kempf+mm}, we give a brief overview of the modified uncertainty relations with minimal
length. We also define the harmonic oscillator model in this framework. We show
that a straightforward factorization method, well familiar from quantum mechanics textbooks,
does not work because the Heisenberg-Weyl algebra of the model is not closed
(it requires an infinite number of operators for closure).
In Section \ref{sec3} we are inspired by the transformation found in Ref.\ \cite{kempf+mm}
that maps the particle momentum into the particle wave-vector and demonstrates explicitly that
plane waves have minimal wavelength.
Using this transformation we calculate the Green's operator of the
harmonic oscillator in the minimal length quantum mechanics and find that it exactly equals to the
Green's operator of the P\"oschl-Teller model. This demonstrates that two systems are equivalent.
Next, as explained in Ref.\ \cite{nieto}, we transform the P\"oschl-Teller model into its natural
coordinates. Described in natural coordinates, the particle moving in the symmetric P\"oschl-Teller potential
appears as if it is exhibiting harmonic oscillation of the ordinary theory but with energy dependent frequency.
It is essential that natural
coordinates are such that the Heisenberg-Weyl algebra is closed.
In Section \ref{sec4} we construct a pair of mutually adjoint ladder operators
for the symmetric P\"oschl-Teller model in its natural coordinates.
We use no knowledge of the exact solution of the model.
The ladder operators for the P\"oschl-Teller model were constructed previously in
Ref.\ \cite{dong+lemus} but the knowledge of the exact solution of the Schr\"odinger equation
was an essential part of the construction. Construction of operators presented in Refs.\ \cite{dong+lemus} and
\cite{meljanac} are essentially identical.
The ladder operators of the symmetric P\"oschl-Teller model satisfy a deformed version of
the Heisenberg-Weyl algebra that also happens to be some particular deformation of $SU(1,1)$ algebra.
We use this deformed algebra to calculate energy eigenstates and eigenvalues exactly.
The energy spectrum and wave-functions agree with previous results.
In Section \ref{sec5} we discuss the physics behind the construction of ladder operators.
In particular we explain where the prominent features such as minimal length
of the deformed quantum mechanics wind up in the framework of the P\"oschl-Teller model.
In Section \ref{sec6} we compare our results with prior works.
We show that the algebra of ladder operators we constructed can
be thought as a deformation of a simple ordinary Bose oscillator algebra or
as a deformation of some $SU(1,1)$, that itself is a deformation of an ordinary Bose oscillator algebra.
We also point out that the dynamical symmetry group of the system is just the dynamical $SU(1,1)$ algebra
of the ordinary Bose oscillator constructed from the quadratic combination of Bose oscillators.
The reason that the dynamical algebra is unchanged is related to the fact that the
deformations do not mix states of different parity.
In Section \ref{sec7} we consider the D-dimensional isotropic harmonic oscillator model \cite{minic-atall}.
We show that it can be analyzed the same way as the one-dimensional model and
that the D-dimensional isotropic harmonic oscillator model in a non-commutative
quantum mechanics with minimal length uncertainty relations is in fact equivalent to a
generalized P\"oschl-Teller model. The appearance of the P\"oschl-Teller potential is related to the
quadratic form of the non-relativistic kinetic energy operator.
Finally, in Section \ref{sec8} we summarize our results, describe possible generalizations and consider
some future directions.

\section{\label{sec2}The Minimal Length Uncertainty Relations in One-Dimension and the Harmonic Oscillator Model}

As described in the introduction, there are number of reasons to
consider modified uncertainty relations in quantum mechanics.
Following \cite{kempf+mm}, we consider a simple
deformation of the basic commutator (\ref{eq:zero}), that implies the existence of minimal length uncertainty.
Let $x$ and $p$ be the position and momentum operators, respectively, and let us assume that they obey the basic commutator
\begin{equation}\label{mod-comm}
\left[ x,p \right]= i \left( 1+\beta p^2\right)~.
\end{equation}
On dimensional grounds, $\sqrt{\beta}$ is measured in units of length.
The ordinary quantum mechanics can be considered as a limit of the deformed theory where
$\beta$ tends to zero.
Formally, operators $x$ and $p$ are hermitian but, as shown in \cite{kempf+mm}, $x$
is not self-adjoint. The operator $x$ cannot be diagonalized but it does have real expectation values.  It also has a one-parameter class of self-adjoint extensions.
To obtain information on the position, the best thing we can do  is to
construct states such that for these states  the uncertainty of the operator $x$ is minimal.
A symmetric operator with minimal uncertainty states and real expectation values is called a fuzzy observable
and the corresponding minimal uncertainty states are some coherent states.

The modified commutator (\ref{mod-comm}) implies a modification of the
uncertainty relations. They are given by
\begin{equation}
\Delta x \Delta p \geq \frac{1}{2} \big\vert \big\langle \left[x,p\right]\big\rangle \big\vert~,
\end{equation}
where  angle bracket $ \big\langle \; \big\rangle$ denotes the expectation value. Uncertainties are defined by the usual expressions.
For an operator $O$ we have,
$\langle O\rangle  = \langle\psi\vert O\vert\psi\rangle$
and $(\Delta O)^2 = \langle\psi\vert (O- \langle O\rangle)^2\vert\psi\rangle =
\langle O^2\rangle - \langle O\rangle^2$.
With the modified basic commutators it follows
\begin{equation}\label{eq:uncertainty-reation}
\Delta x \Delta p \geq \frac{1}{2}\left( 1+ \beta \Delta p^2 + \beta \langle p\rangle^2\right)\ .
\end{equation}
The uncertainty relation (\ref{eq:uncertainty-reation})
is saturated when the two sides are equal. The quadratic term present on the right hand side
implies that there exists a minimal uncertainty in the position.
The smallest possible uncertainty in the position occurs for sates that have zero average momentum, $\langle p \rangle =0$. Then,
\begin{equation}
\Delta x_{\mathrm{min}} = \sqrt{\beta}~.
\end{equation}
We will explore the implications of the presence of absolutely the smallest possible
resolution of distance within the context of a harmonic oscillator model.

The harmonic oscillator is probably the most widely studied and used example in all
of physics, quantum and classical. It is a feature-rich model, still simple enough that it
can be solved exactly by a variety of methods.
The fact that it is also a model of the $0+1$ dimensional field theory makes  it
even more attractive in the present context.
It is defined by the Hamiltonian
\begin{equation}\label{eq:ho}
H = \frac{1}{2} p^2 + \frac{\omega^2}{2} x^2\ .
\end{equation}
In the ordinary quantum mechanics, the spectrum and states can be constructed, for example, by representing the Heisenberg-Weyl algebra using a pair of operators, $a$ and
$a^\dagger$, that satisfy the commutator relation
$[a,a^\dagger]=1$. This commutator relation is
equivalent to the basic commutator $[x,p]=i$.
Then, $[H,a] = -\omega a$ and
$[H,a^\dagger] = \omega a^\dagger$. The spectral energies are given by $E_n = \omega(n+1/2)$ where $n=0,1,\ldots,$ and states are constructed as
$\vert n\rangle \propto ({a^\dagger})^n\vert 0\rangle$.
The ground state is defined by $a\vert0\rangle =0$. Furthermore, the states are classified
by the  dynamical $SU(1,1)$ symmetry algebra constructed as follows. Let $S_{+}= \frac{1}{2} {a^\dagger}^2$,
$S_{-}= S^\dagger_+$ and $S_{0} = \frac{1}{4}(a a^\dagger + a^\dagger a)$. Operators $S_\pm$ and
$S_0$ satisfy the algebra, $[S_+,S_-]= -2S_0$ and $[S_0,S_\pm] = \pm S_\pm$.
The Hamiltonian is given by $H= 2\omega S_0$. The spectrum is characterized by a Bargman
index $k$ which determines the quadratic Casimir operator $C$ of the dynamical $SU(1,1)$
algebra. We have $C = S_0^2 - \frac{1}{2}(S_+S_-+S_-S_+) = k(k-1)$.
For the ordinary oscillator $C = -3/16$ and it
corresponds to $k = 1/4$ and $k=3/4$.
The space of states of the oscillator forms a reducible representation of the dynamical symmetry group
and splits into two subspaces  each forming an infinite-dimensional,  irreducible representation of $SU(1,1)$.
Even parity oscillator sates correspond to Bargman index $k=1/4$  and odd parity oscillator states correspond to Bargman index $k=3/4$.

This construction and characterization of the oscillator states is possible due to three facts:
\begin{itemize}
\item[1)] The states $\vert n\rangle \propto (a^\dagger)^n\vert 0\rangle$ form a basis of
the Hilbert space on which the commutator $[a,a^\dagger]$ is diagonal.
\item[2)] The Heisenberg-Weyl algebra of the oscillator, given by commutators
$[H,x] = -ip$ and $[H,p] = i\omega^2 x$ is closed. Closure of the algebra means
that multiple commutators
involving the Hamiltonian do not involve any new operators in addition to $x$ and $p$.
\item[3)] Operators $x$ and $p$ and operators $a$ and $a^\dagger$ are related by a linear transformation.
\end{itemize}
A few  comments are in order here. In the ordinary
quantum mechanics the basic commutator  $[x,p]=i$ is diagonal so
it is always possible to choose $[a,a^\dagger]$ to be diagonal.
This means that 1) can be trivially satisfied.
It is sufficient that operators $a$ and $a^\dagger$ in 1) be related to operators
$a$ and $a^\dagger$ in 3) by a $SU(1,1)$ transformation.
The $SU(1,1)$ here is the dynamical symmetry of the model and it is realized linearly.
In the case of the harmonic oscillator the commutator in 1) takes the
simplest possible form, because it is a unit operator. This is achieved by using the simple
factorization based on the transformation
$x = \frac{1}{\sqrt{2\omega}} (a + a^\dagger)$,
$p = -i\sqrt{\frac{\omega}{2}}(a - a^\dagger)$.

\medskip

We would like to arrive at the description of the oscillator in the quantum theory 
with minimal uncertainty relations that
is  parallel to that of the oscillator in ordinary quantum mechanics. As we show in this paper, this is
possible, but highly nontrivial.

The quantization of the system is essentially equivalent to  finding a basis set in the Hilbert space that simultaneously
diagonalizes the commutator and the Hamiltonian.
In the ordinary quantum mechanics the basic commutator is diagonal in any basis because it
is proportional to a unit operator, Eq. (\ref{eq:zero}).
Hence, any basis that diagonalizes the Hamiltonian will do.
In the deformed quantum mechanics this is no longer the case.

The deformation of the basic commutator has a profound effect on the Heisenberg-Weyl algebra. Consider two
commutators, $[H,x] = -ip - i\beta p^3$ and
$[H,p]=i\omega^2x +\omega^2\beta p +i\omega^2\beta xp^2 + \omega^2\beta^2 p^3$, valid for the
oscillator in the deformed quantum mechanics.
Clearly, this algebra is not a closed algebra. In addition to operators $x$ and $p$, new operators that
are quadratic and  cubic in $x$ and $p$ appear. These higher powers of operators appear in the algebra
precisely because of the modification of the basic commutator in Eq.\ (\ref{mod-comm}).
Computing additional commutators such as
$[H,[H,x]]$ and $[H,[H,p]]$ we
find that higher an higher powers of basic operators enter. This simply means that
the Heisenberg-Weyl algebra of the model is not closed. This makes the application of the
operator factorization method highly nontrivial.

\section{\label{sec3}Equivalence of the One-Dimensional Harmonic Oscillator With Minimal Length
Uncertainty Relation and the P\"oschl-Teller Model}

The modified basic commutator (\ref{mod-comm}) implies that there is minimal length below which it is
not possible to resolve distances. The free particle states are still plane waves but they exhibit minimal
wavelength, $\lambda_{min} = 4\sqrt{\beta}$. The dispersion relation of the free particle of mass $m$
found in Ref.\ \cite{kempf+mm} is given by
\begin{equation}\label{eq:dispersion}
E_{\mathrm{free~particle}} = \frac{1}{2m\beta} \left(\tan \frac{2\pi\sqrt{\beta}}{\lambda} \right)^2~.
\end{equation}
The transformation that maps the free particle energy eigenvalue $E= \frac{1}{2m}p^2$ into the wave-vector
representation is given by
\begin{equation}\label{eq:transf}
p = \frac{1}{\sqrt{\beta}}\tan \sqrt{\beta}\rho~,
\end{equation}
where $\rho = 2\pi/\lambda$ is the particle wave-vector.

It was also shown in Ref.\ \cite{kempf+mm} that the commutator relation ({\ref{mod-comm})
can be realized in the momentum representation by taking the position operator in the momentum representation as
\begin{equation}\label{momentum-rep}
x = \left(1+\beta p^2\right)\left(i\frac{d}{dp}\right) +i\gamma p~.
\end{equation}
The momentum operator $p$ acts simply as multiplication, and
 $\gamma$ is some parameter that can be chosen freely \cite{minic-atall}.
Operators $x$ and $p$ are formally hermitian with respect to a measure
\begin{equation}
d\mu(p) = \frac{dp}{(1+\beta p^2)^{1-\gamma/\beta}}~
\end{equation}
on the  ($-\infty \leq p\leq \infty$) interval.
In Refs.\ \cite{kempf+mm} and  \cite{minic-atall} the representation given by Eq. (\ref{momentum-rep}) was used to formulate and solve the Schr\"odinger equation for the problem,
\begin{equation}
H\psi(p) = E\psi(p)~.
\end{equation}
They found that the states are labeled by a single quantum number $n=0,1,\ldots,~$. The wave functions $\psi(p)$
are essentially given by Gegenbauer polynomials and $E$ is a quadratic function of $n$.

In momentum space representation the kinetic energy in the Schr\"odinger equation appears as some kind of potential energy term. In fact, Eqs.\ (\ref{eq:transf}) and (\ref{eq:dispersion}) are very suggestive.
The kinetic energy term in the Schr\"odinger equation has an appearance of the potential of the symmetric
P\"oschl-Teller model. To exploit this relationship we calculate the Green's function $G(z)$ of the
oscillator and show that it equals exactly to the Green's function of the P\"oschl-Teller model. This establishes
the the equivalence of the two models.

Let $z$ be a complex number. The Green's function is given formally by
\begin{equation}
G(z) = \left( z-H\right)^{-1}~.
\end{equation}
Let $\Psi(p)$ and $\Phi(p)$ be two arbitrary state functions of the oscillator in the deformed quantum mechanics.
We now calculate the matrix element of $G^{-1}(z)$
\begin{equation}
\begin{split}
\langle\Phi\vert &  G^{-1}(z)\vert\Psi\rangle  \\
= & \int_{-\infty}^\infty d\mu(p)~ \Phi^\ast(p)
\left[ z - \frac{1}{2}\left( p^2 +\omega^2 x^2\right)
\right]\Psi(p)~.\\
\end{split}
\end{equation}
Making first the variable change given by Eq.\ (\ref{eq:transf})
and then performing a similarity transformation that removes the parameter $\gamma$,
$\Psi = J\psi$ with
$J = (\cos\sqrt{\beta}\rho)^{\gamma/\beta}$, we obtain
\begin{equation}
\langle\Phi\vert G^{-1}(z)\vert\Psi\rangle =
\int\limits_{-\pi/2\sqrt{\beta}}^{\pi/2\sqrt{\beta}}
d\rho~ \phi^\ast(\rho)\left( z - H^\prime\right)\psi(\rho)~,
\end{equation}
where
\begin{equation}\label{SPT-nieto}
\begin{split}
H^\prime\left(i\frac{d}{d\rho},\rho \right) = &
J^{-1} H\left(i\frac{d}{d\rho},\rho\right) J =  \\
& -\frac{\omega^2}{2}\frac{d^2}{d\rho^2} +
\frac{1}{2\beta} \tan^2\sqrt{\beta}\rho~.\\
\end{split}
\end{equation}
The Hamiltonian $H^\prime$ is the Hamiltonian of the symmetric P\"oschl-Teller
model.
We note that the deformed commutator (\ref{mod-comm}) becomes simply the commutator of the ordinary
quantum mechanics.
\begin{equation}\label{new-comm}
\left[i\frac{d}{d\rho},\rho \right] = i~.
\end{equation}
We can bring Hamiltonian $H^\prime$
to a standard form of P\"oschl-Teller model by rescaling a variable and defining new constants.
Define
\begin{equation}\label{eq:new-constants}
\begin{split}
& \sqrt{\beta}\rho = \alpha x~,~~~~\alpha^2 = \omega^2\beta \\
& \frac{1}{\beta} = \frac{\omega^2}{\alpha^2} = \alpha^2 \nu(\nu -1)~\\
\end{split}
\end{equation}
and the measure becomes
$\int\limits_{-\pi/2\sqrt{\beta}}^{\pi/2\sqrt{\beta}} d\rho = \frac{\alpha}{\sqrt{\beta}}
\int\limits_{-\pi/2\alpha}^{\pi/2\alpha} dx$.
We also introduce $p = -i\frac{d}{dx}$ such that $[x,p]=i$ -- this is
compatible with Eq.\  (\ref{new-comm}).
Then the Hamiltonian reads
\begin{equation}
\begin{split}
H^\prime (p,x) & = \frac{1}{2} p^2 + \frac{\alpha^2}{2}
\frac{\nu(\nu -1)}{\cos^2\alpha x} - \frac{1}{2\beta} \\
& = H_{\mathrm{SPT}}(x,p) - \frac{1}{2\beta}~,
\end{split} \end{equation}
where $H_{\mathrm{SPT}}$ is the Hamiltonian of the symmetric P\"oschl-Teller model in standard form.

The symmetric P\"oschl-Teller model has been extensively studied before.
It is exactly solvable by a variety of methods.
In Ref.\ \cite{dong+lemus} it was shown that its Heisenberg-Weyl algebra is a deformed $SU(1,1)$;
see also \cite{daskaloyannis}.

In Ref.\ \cite{nieto}, it was shown that a very general potential that supports bound states can always be
reformulated in terms of some so-called natural variables with the property
that motion looks like the motion in the harmonic oscillator potential with energy dependent
frequency. In these natural coordinates coherent states considered in \cite{nieto}
obey essentially a classical equation of motion of the harmonic oscillator.
These natural coordinates no longer satisfy canonical commutator relations of the
ordinary quantum mechanics. However, the basic commutator between the natural coordinate and the conjugate momenta
does not imply any limit on uncertainties because its character is different from
deformation given by Eq. (\ref{mod-comm}). The Heisenberg-Weyl algebra of the model expressed
in natural coordinates is essentially the Heisenberg-Weyl algebra of oscillator in ordinary quantum mechanics and it is closed.

The natural coordinates for the P\"oschel-Teller model used in \cite{nieto} are defined as follows:
\begin{equation}\begin{split}
& y = \sin\alpha x~,\\
&k = \frac{\alpha}{2}\{ \cos\alpha x,p\} = \alpha\cos\alpha x~ p +i\frac{\alpha^2}{2}\sin\alpha x~.\\
\end{split}
\end{equation}
In natural coordinates the matrix elements of the operator $G^{-1}(z)$ is given by
\begin{equation}
\begin{split}
\langle & \Phi\vert G^{-1}(z)\vert\Psi\rangle  \\
=&  \frac{1}{\sqrt{\beta}}
\int\limits_{-1}^1 d\mu(y)~\phi^\ast(y)\left[ z^\prime - H_\mathrm{SPT}(k,y)\right]\psi(y)~,
\end{split}
\end{equation}
where the new measure is $d\mu(y) = \frac{dy}{\sqrt{1-y^2}}$.
We have also made a  shift of the energy variable,
$z^\prime = z +\frac{1}{2\beta}$.
In natural variables the Hamiltonian reads
\begin{equation}\label{SPT-yk}
\begin{split}
H_{\mathrm{SPT}}(k,y) = & \frac{1}{2\alpha^2}
\left(
\frac{1}{\sqrt{1-y^2}}
\left( k -i\frac{\alpha^2}{2} \right)
\right)^2 \\
+ &
\frac{\alpha^2}{2}\frac{\nu (\nu -1)}{1-y^2}\ .\\
\end{split}
\end{equation}
In the next section we will show that the Heisenberg-Weyl algebra of the symmetric P\"oschl-Teller model
in natural coordinates is a closed algebra and that it can be used to  construct states and spectral energies
algebraically.

\section{\label{sec4}The Heisenberg-Weyl Algebra Of The Symmetric P\"oschl-Teller Model}

In this section we construct the spectral algebra for the Hamiltonian
$H_{\mathrm{SPT}}(k,y)$, Eq. (\ref{SPT-yk}).
A straightforward calculation yields
(for the ease of writing we drop the subscript on $H_{\mathrm{SPT}}$ in what follows):
\begin{equation}\label{spt-alg-space}
\begin{split}
\left[ y,k\right] = & i\alpha^2 (1-y^2)\\
\left[ H , y\right]  = & -ik \\
\left[ H, k\right] = & i\alpha^2\left(2yH -\frac{\alpha^2}{4} y - ik \right)~.
\end{split}
\end{equation}
Calculating the commutators $[H,[H,y]]$ and $[H,[H,k]]$ we find that no new operators are generated.
Hence the algebra is closed. This means that there exist combinations of
operators $y$ and $k$ that can serve as spectral operators.

Note that the right-hand side of Eq.\ (\ref{spt-alg-space}) depends on the Hamiltonian $H$ reflecting
the energy dependence of oscillating frequency.
This also means that the resulting algebra is deformed.
To find the correct combination of operators $y$ and $k$,  that serve as spectral operators, is nontrivial.
The essence of the structure of ladder operators can be guessed from the work in Ref. \cite{nieto}. 
We can formalize the calculation  as follows. 
Note that the last two equations of (\ref{spt-alg-space}) can be written in a matrix form,
\begin{equation}\label{eq:18}
H \left(\begin{matrix} y & k \\ \end{matrix} \right) =
\left(\begin{matrix} y & k \\ \end{matrix} \right)
\left(
\begin{matrix}
H & i\alpha^2 (2H- \alpha^2/4) \\
-i & H+\alpha^2 \\
\end{matrix}
\right)~.
\end{equation}
Matrix on the right hand side can be diagnoalized by a simple similarity transformation,
$M= JM_dJ^{-1}$ where
\begin{equation}
J = \left(
\begin{matrix}
-i\alpha (\sqrt{2H} +\alpha/2) & i\alpha (\sqrt{2H} -\alpha/2) \\
1 & 1 \\
\end{matrix}
\right)~.
\end{equation}
The diagonal matrix $M_d$ reads,
\begin{equation}
M_d=
\left(
\begin{matrix}
H+ \alpha^2/2 - \alpha\sqrt{2H} & 0 \\
0 & H+ \alpha^2/2 + \alpha\sqrt{2H} \\
\end{matrix}
\right)~.
\end{equation}
The spectral operators are essentially two combinations given by
$ \left( \begin{matrix} y & k \\ \end{matrix} \right)J$.
It follows immediately that the following two operators can serve as spectral ladder operators:
\begin{equation}\label{cr-ann-op-def}
\begin{split}
a = & \frac{1}{\alpha^2}\left[y \left(\alpha\sqrt{2H}+ \frac{\alpha^2}{2}\right)+ik\right]\\
a^\dagger = & \frac{1}{\alpha^2}\left[\left(\alpha\sqrt{2H}+\frac{\alpha^2}{2}\right)y-ik\right]~.\\
\end{split}
\end{equation}

The operators $a$ and $a^\dagger$ obey an algebra that can be used to construct
the spectrum of the system,
\begin{equation}\label{spectral-alg-1}
\begin{split}
& \left[ H, a\right] = - a \left( \alpha\sqrt{2H} - \frac{\alpha^2}{2}\right) =
- \left(\alpha\sqrt{2H} + \frac{\alpha^2}{2} \right) a\\
&\left[ H, a^\dagger\right] =   \left( \alpha\sqrt{2H} - \frac{\alpha^2}{2}\right) a^\dagger =
a^\dagger \left(\alpha\sqrt{2H} + \frac{\alpha^2}{2} \right)~.\\
\end{split}
\end{equation}
An important side benefit of this construction is that we can also evaluate any commutator of the
form $[F(H),y]$ and $[F(H),k]$ where $F(H)$ is an analytic function of the Hamiltonian. We simply
read of the needed relations from the matrix equation
$ F(H) \left( \begin{matrix} y & k \\ \end{matrix} \right)J =
\left(\begin{matrix} y & k \\ \end{matrix}\right)J F(M_d) $.
Alternatively, we can also use this result to evaluate
any commutators of the form $[F(H),a]$ and $[F(H),a^\dagger]$. In fact, the second set of equalities in
(\ref{spectral-alg-1}) was determined this way.

The calculation of the commutator and the anti-commutator of
$a$ and $a^\dagger$ is straightforward but  lengthy.
It is the best to calculate the following two products first
\begin{equation}\label{aadagger}
\begin{split}
aa^\dagger = & -\nu(\nu -1)\frac{\frac{\sqrt{2H}}{\alpha}+1}{\frac{\sqrt{2H}}{\alpha}} +
\left(\frac{\sqrt{2H}}{\alpha} + 1 \right)^2\\
a^\dagger a = & -\nu(\nu -1) \frac{\frac{\sqrt{2H}}{\alpha}}{\frac{\sqrt{2H}}{\alpha} - 1 }+
\left(\frac{\sqrt{2H}}{\alpha}\right)^2~. \\
\end{split}
\end{equation}
Then we obtain
\begin{equation}\label{algebra-rest}
\begin{split}
\left[ a,a^\dagger\right] = & f(H) \\
f(H) = &  1+ 2 \frac{\sqrt{2H}}{\alpha} +
\frac{\nu (\nu -1)}{\frac{\sqrt{2H}}{\alpha}(\frac{\sqrt{2H}}{\alpha}-1)} \\
\left\{ a,a^\dagger\right\} = & \frac{1}{2}
-\nu(\nu -1) \frac{2\left(\frac{\sqrt{2H}}{\alpha}\right)^2 - 1}
{\frac{\sqrt{2H}}{\alpha}(\frac{\sqrt{2H}}{\alpha}-1)} \\
+ &  \frac{1}{2}\left(2 \frac{\sqrt{2H}}{\alpha}+ 1\right)^2 ~. \\
\end{split}
\end{equation}
The algebra also has a quadratic invariant Casimir operator $C$, \cite{polychronakos} and \cite{silvio}.
It equals $C = \nu(\nu-1)$ and it codes the strength of the potential. 

The spectral algebra of the model is given by Eqs.\ (\ref{spectral-alg-1}) and  (\ref{algebra-rest}).
It is a two-function deformed $SU(1,1)$ algebra. If we make the identifications
\begin{equation}
a^\dagger \leftrightarrow S_+~,~~~~~~
a \leftrightarrow S_-~,~~~~~~
H \leftrightarrow S_0~,
\end{equation}
we have
\begin{equation}
\begin{split}
\left[ S_+,S_-\right] = & f(S_0)\\
\left[ S_0,S_+\right]= & g(S_0) S_+\\
\left[ S_0,S_-\right] = & - S_-g(S_0)~,\\
\end{split}
\end{equation}
where $g(S_0) = \alpha \sqrt{2S_0} - \frac{\alpha^2}{2}$.

Now we show that the complete spectrum of the system can be constructed from the algebra alone.
A representation is characterized by a parameter, see Eq.\ (\ref{eq:new-constants}),
\begin{equation}
\nu = \frac{1}{2}\left(1+\sqrt{1+\frac{4}{\beta^2\omega^2}}\right)~.
\end{equation}
The ground state is defined by
\begin{equation}\label{ground-state}
a \vert \psi_{0};\nu\rangle = 0~,~~~~~~
H\vert \psi_{0};\nu\rangle = E_0 \vert \psi_{0},\nu\rangle~.
\end{equation}
Using Eq.~(\ref{cr-ann-op-def}) we can convert this relation into a first order
differential equation and we can obtain the ground state wave-function in $y$-space representation
\begin{equation}
\psi_0(y) = \langle y\vert \psi_{0};\nu\rangle =
\sqrt{\frac{\alpha\Gamma\left(\nu +1\right)}{\sqrt{\pi}\Gamma\left(\nu +\frac{1}{2}\right)}}
(1-y^2)^{\nu/2}~.
\end{equation}
The easiest way to determine the ground state energy is to evaluate the
expectation value of $a^\dagger a$ in the ground state.
Then, we get
\begin{equation}
E_0 =  \frac{\alpha^2\nu^2}{2}~.
\end{equation}

From the spectral algebra (\ref{spectral-alg-1}) it is clear that operators $a$ and $a^\dagger$
act as energy-state lowering and raising operators, respectively.
Excited states are obtained by applying powers of the creation operator $a^\dagger$ on the ground state,
\begin{equation}\label{spt-state}
\vert \psi_n;\nu\rangle = N_n \left( a^\dagger\right)^n\vert\psi_{0};\nu\rangle~,
\end{equation}
where $N_n$ is a normalization constant.

Let the state $\vert\psi_n;\nu\rangle$ be an eigenstate of the Hamiltonian $H$
with energy $E_n$.
Then the state $a^\dagger\vert\psi_n;\nu\rangle$ is also an eigenstate
of $H$ but with energy $E_{n+1}$ and the state $a\vert\psi_n;\nu\rangle$ is an eigenstate of $H$
with energy $E_{n-1}$:
\begin{equation}\label{spectra-1}
\begin{split}
 H\vert\psi_n;\nu\rangle & =  E_n\vert\psi_n;\rangle \\
 Ha^\dagger \vert\psi_n;\nu\rangle  & =
 a^\dagger\left(H+\alpha \sqrt{2H} +\alpha^2/2 \right)\vert\psi_n;\nu\rangle\\
 & = \left( E_n + \alpha\sqrt{2E_n} +\alpha^2/2 \right) a^\dagger\vert\psi_n;\nu\rangle \\
 & = E_{n+1}\vert\psi_n;\nu\rangle \\
 Ha \vert\psi_n;\nu\rangle  & = a\left( H - \alpha\sqrt{2H} +\alpha^2/2\right)\vert\psi_n;\nu\rangle \\
& =  \left( E_n -\alpha\sqrt{2E_n} +\alpha^2/2\right) a \vert\psi_n;\nu\rangle  = \\
& = E_{n-1} a\vert \psi_n;\nu\rangle ~. \\
\end{split}
\end{equation}
Equation (\ref{spectra-1}) can be rearranged to read
\begin{equation}\label{energy-1}
\sqrt{2E_{n+1}} = \sqrt{2E_n}+\alpha~.
\end{equation}
By iterating this relation, starting from the ground-state energy, we get the energy spectrum.
The result for the P\"oschl-Teller model reads
\begin{equation}
E_n^{\mathrm{SPT}} = \frac{\alpha^2}{2} (n+\nu)^2~,~~~~~~ n = 0,1,2,\ldots,
\end{equation}
and it is in agreement with the well known result \cite{haar}.

Recall that the energy of the oscillator with minimal length uncertainty relations is shifted relative to that of
the symmetric P\"oschl-Teller model. Therefore
\begin{equation}
\begin{split}
& E^{\mathrm{OSC}}_n = \frac{\alpha^2}{2}( n+\nu)^2 -\frac{1}{2\beta}= \\
& \omega \left(n+\frac{1}{2}\right)\sqrt{1+\frac{\omega^2\beta^2}{4}} +
\frac{\omega^2\beta}{2}\left(\left(n+\frac{1}{2}\right)^2 +\frac{1}{4}\right)~.
\end{split}
\end{equation}
This is also in agreement with the previous results in \cite{kempf+mm} and \cite{minic-atall}.

It is not difficult to derive the explicit relations between states $\vert\psi_n;\nu\rangle$ and states
$\vert\psi_{n\pm 1};\nu\rangle$. We have
\begin{equation}
a^\dagger\vert\psi_n\rangle = \kappa_{n+1}\vert\psi_{n+1}\rangle~,~~~~~~
a\vert\psi_n\rangle = \kappa_n\vert\psi_{n-1}\rangle~.
\end{equation}
Taking a diagonal matrix element of $[a,a^\dagger]=f(H)$ and using the explicit expression for energy eigenvalue
we obtain the recursion relation for the coefficients
$\kappa_n$
\[
\begin{split}
& \langle \psi_n;\nu\vert\left[ a,a^\dagger\right]\vert \psi_n;\nu\rangle =
\vert\kappa_{n+1}\vert^2 - \vert\kappa_n\vert^2 = f(E_n)\\
& = 1+2(n+\nu) +
\frac{\nu(\nu -1)}{(n+\nu)(n+\nu -1)}~.\\
\end{split}\]
It is easy to find the solution (we take $\kappa_n$ to be real)
\begin{equation}
\begin{split}
\kappa_n & = \sqrt{(n+\nu)^2-\nu(\nu -1)\frac{n+\nu}{n-1+\nu}} \\
& = \sqrt{\frac{n+\nu}{n-1+\nu}} \sqrt{n(n+2\nu -1)}~.
\end{split}
\end{equation}
The normalization constant $N_{n}$  can also be computed. From Eq.\ (\ref{spt-state}),
we have
\begin{equation}\label{for-recursion}
\begin{split}
\vert\psi_n;\nu\rangle = & N_n (a^\dagger)^n\vert 0;\nu\rangle \\
& \frac{N_n}{N_{n-1}} a^\dagger\vert\psi_{n-1};\nu\rangle =
\frac{N_n}{N_{n-1}}\kappa_n\vert\psi_n;\nu\rangle~.
\end{split}
\end{equation}
Then, using $\langle 0;\nu\vert 0;\nu\rangle = 1$, and iterating, we calculate
\begin{equation}
N_n = \frac{1}{\kappa_n\kappa_{n-1}\ldots \kappa_1} =
\sqrt{\frac{\nu \Gamma(2\nu)}{(\nu+n)n!\Gamma(2\nu +n)}}~.
\end{equation}
It is not too difficult to calculate the wave function in the $y$-space. Let
\begin{equation}
\psi_n^{\nu}(y) = \langle y\vert\psi_n;\nu\rangle~.
\end{equation}
From Eq. (\ref{for-recursion}), using the creation operator written in terms of $k$ and $y$ operators,
Eq. (\ref{cr-ann-op-def}), we find
\begin{equation}
\begin{split}
\psi_n^\nu(y) = & \frac{1}{\alpha^2\kappa_n}
\left(y g(E_{n-1}) -\alpha^2 (1-y^2)\frac{d}{dy}  \right. \\
& \left. +\frac{\alpha^2}{2} y\right)\frac{\sqrt{2E_{n-1}}+\alpha}{\sqrt{2E_{n-1}}}\psi_{n-1}^\nu(y)~. \\
\end{split}
\end{equation}
Starting from the explicit wave function for the ground state  and making use of the
differential equation satisfied by Gegenbauer polynomials, see Ref.\ \cite{gradshteyn}, we  find
\begin{equation}\label{wave-function-gegenbuer}
\psi_n^\nu(y) = 2^\nu \Gamma(\nu)\sqrt{\frac{\alpha n! (n+\nu)}{2\pi\Gamma(n+\nu)}}
(1-y^2)^{\nu/2} C_n^\nu(y)~,
\end{equation}
where $C_n^\nu(y)$ is a Gegenbauer polynomial.
From  this we can obtain immediately the wave function of an oscillator in
the  minimal length quantum mechanics \cite{minic-atall}
\begin{equation}
\Psi^{\mathrm{OSC}}_n(y) = 2^\nu \Gamma(\nu)\sqrt{\frac{\alpha n! (n+\nu)}{2\pi\Gamma(n+\nu)}}
(1-y^2)^{\left(\frac{\nu+\gamma/\beta}{2}\right)} C_n^\nu(y)~.
\end{equation}

In this section we have constructed, by using algebraic factorization methods,
and without the explicit knowledge of the exact solution,
a pair of creation and annihilation operators for the model. The two operators obey a deformed
$SU(1,1)$ algebra. Then we have calculated the exact
energy eigenvalues and the energy eigenstates. The states are characterized by a single parameter $\nu$
that is determined by the strength of the potential in the symmetric P\"oschl-Teller model case or
by the parameter $\beta$ that measures deformation of the uncertainty relation of the quantum mechanics and gives the fundamental, minimal possible resolution of length.

\section{\label{sec5}The Physics Behind the Construction}

In this section we offer some insight into the problem of quantization in the deformed quantum mechanics.

The deformed commutator (\ref{mod-comm})  can be written as
\begin{equation}\label{eq:48}
[x,p] = i (1+ 2m\beta H_0)~,~~~~~~
H_0 = \frac{1}{2m} p^2~,
\end{equation}
where $H_0$ is the Hamiltonian of a free particle of mass $m$.
The Heisenberg-Weyl algebra of the oscillator reads
\begin{equation}\label{eq:50}
\begin{split}
[ H, x ] = & -ip (1+2m\beta H_0) \\
[ H, p ] = & i\omega^2  x (1+2m\beta H_0) + \omega^2 m\beta p (1+2m\beta H_0)~. \\
\end{split}
\end{equation}
However, these equations are incomplete because there are, in fact, four relevant operators,
$x,p,H_0$ and $H$, at the start. We need additional commutators
\begin{equation}\label{eq:51}
\begin{split}
[H_0,p] = & 0 \\
[H_0,x] = &  -i p (1+2\beta H_0)\\
[H,H_0] = & i\frac{\omega^2}{2} (xp+px) (1+2\beta H_0) \\
 +& 2\beta \omega^2 H_0 (1+2\beta H_0)~.\\
\end{split}
\end{equation}
Eqs.\ (\ref{eq:48}), (\ref{eq:50}) and (\ref{eq:51}) clearly highlight the source of the problem.
In the deformed theory, the algebra contains two Hamilton operators.
Together, they fail to close the algebra. Their commutator generates new operators such as
$xp+px$ and $H_0^2$. We can try to modify the algebra from the start by adding all these new operators
but this is of no help. New operators when added will generate more new operators and cycle will never end.

Note however, that $H_0, p$ and $x$ form a closed subalgebra.
This offers a possibility to find the quantum theory of some $H_0^\prime \propto H_0$ in
$H_0$ deformed quantum theory. This will select a basis in the Hilbert space  such that the
basic commutator $[x,p]$ is diagonal. This basis can then be used to compute the matrix elements of $H$.
Of course, in this basis $H$ is not diagonal.
The remaining problem then is to find the unitary transformation $U$ in the Hilbert space such that
$H^\prime = U^\dagger HU$ is diagonal. In general, this can be difficult.
Final states that diagonalize $H$ are linear combinations of
states that diagonalize $H_0$. Possibly, these are some coherent states.
As a result, solving the quantum problem in the deformed theory doubles the work as we have to solve more than just one quantization problem. The reason why the ordinary quantum theory is much easier is simply a
consequence of the fact that in the ordinary quantum theory the basic commutator is already diagonal and
requires no additional work.

In fact, the two transformations we used in Sections \ref{sec3} and \ref{sec4} to quantize the oscillator
\begin{equation}
p = \frac{1}{\sqrt{\beta}} \tan\sqrt{\beta}\rho~,~~~~~~\&~~~~~~
y = \sin\sqrt{\beta}\rho~,
\end{equation}
carry out the program we just described.
For another possibility see \cite{spector} and \footnote{\label{foot1}
We should note
that in principle we can start
with $H$ in place of $H_0$. Of course, this changes the physics drastically because the basic commutator is now given as  $[x,p] = i(1+2\gamma H)$, ($\gamma$ is some scale constant).
The basic commutator with $H$ implies the existence of minimal momentum in addition to the
minimal length. However, it may be possible that this new problem is easier to solve.
In the new theory we first diagonalize $H$ with itself as a deformation. Then we
quantize $H$ again but with altered parameters.
Perhaps the shape invariance of the supersymmetric quantum mechanics can be useful in this case,
\cite{spector}. Then, we can carry out the limit that removes minimal momentum uncertainty
to recover the minimal length uncertainty theory.}.

In the present case there are also some lucky circumstances related to the fact that the potential energy of the harmonic oscillator is simply a quadratic operator. Any other potential energy function would be
more complicated. The  situation then reminds us of the case of the Klein-Gordon relativistic equation where
the Hamiltonian is given by $H = \sqrt{c^2 p^2 + m^2 c^4}$ and $c$ stands for the speed of the light.
The solution is to formulate the equation based on $H^2$, or to linearize with the price of introducing
multi-component wave-functions. It is not clear at present how far one can carry out such enterprize
for interesting potentials. Another possibility would be to expand the more complicated
potential energy $V(x)$ around the oscillator but this is likely to run into convergence issues.
Perhaps the answer is to study more closely the natural coordinates of \cite{nieto} for other potentials.

\section{\label{sec6}Comparison With  Prior Work}

In this section we show the connection of our results with prior works.

We begin by taking a closer look at the Heisenberg-Weyl algebra
realized in terms of  operators $a$ and $a^\dagger$ in Eq.~(\ref{spectral-alg-1}).
We observe that it can be written in a simpler form that closely resembles
the algebra of the harmonic oscillator in the ordinary quantum mechanics. Define a new operator
\begin{equation}
N = \frac{1}{\alpha}\sqrt{2H} -c~,
\end{equation}
where $c$ is some constant we will fix shortly.
The part of the algebra involving the Hamiltonian takes on the
form satisfied by the number operator,
\begin{equation}
[N, a^\dagger] =   a^\dagger~,~~~~~~~~
[N , a] =  -  a~.
\end{equation}
However, this is still a deformed algebra and not the algebra of the oscillator
in the ordinary quantum mechanics because the commutator of $a$ and $a^\dagger$ is deformed
\begin{equation}\label{eq:56}
\begin{split}
[ a,a^\dagger] = & 1+ 2(N+c)  + \frac{\nu(\nu-1)}{(N +c)(N +c -1)} \\
= & \phi(N+1) - \phi(N)~, \\
\end{split}
\end{equation}
where
\begin{equation}\label{eq:phi}
\begin{split}
a^\dagger a = & \phi(N) = (N+c)^2 -  \frac{\nu(\nu-1)}{N+c-1} - \nu(\nu-1) \\
a a^\dagger = & \phi(N+1)~. \\
\end{split}
\end{equation}
Using Eq.\ (\ref{eq:phi}) we can understand the spectrum a bit better.
Let $\vert 0\rangle$ be a normalized ground state defined by
\begin{equation}
a\vert 0\rangle = 0~.
\end{equation}
We want to interpret operator $N$ as a number operator.
This means that we expect $N\vert 0\rangle=0$.
It then follows,
$\phi(N)\vert 0\rangle =  \phi(0)\vert 0\rangle = 0$, or
\begin{equation}\label{eq:phi-condition}
\phi(0)=0~.
\end{equation}
This equation determines the
constant $c$.
There are three possible solutions, $c=0, \nu$ and $1-\nu$. The solution $c=0$ is not acceptable because it implies $\phi_{c=0}(1)=\infty.$
The solution $c=1-\nu$, at least on the surface, appears to be equivalent to the solution
$c=\nu$, because it amounts to a redefinition of the parameterization of the strength
of P\"oschl-Teller potential, see Eq. (\ref{eq:new-constants}).
We choose the solution
\begin{equation}
c=\nu~.
\end{equation}
With this choice the function
\begin{equation}
\phi(n) = (n+\nu)^2 - \frac{\nu(\nu-1)}{n+\nu -1} - \nu(\nu-1)
\end{equation}
has no zeros for any positive integer\footnote{In fact it is possible for the function $\phi(n)=0$ to have a zero
for a special value $\nu=\frac{1-k}{2}$ where $k$ is a fixed integer. In that case the state $\vert k\rangle$
would be zero norm state and we would have the finite dimensional representation.
It is interesting to note that this
possibility is not incompatible with minimal length uncertainty assumption. Using
Eq.~(\ref{eq:new-constants}) we find that in that case $\beta^{-1}= \frac{\omega}{2}\sqrt{k^2-1}$.
This is an intriguing  possibility that implies quantization of parameters $\beta$ or $\omega$, or both.
This situation deserves more study.}.
There is an infinite tower of states of the form
$\vert n\rangle \propto {a^\dagger}^n\vert 0\rangle$.
These are precisely the states we have constructed in Section \ref{sec4}.

We can now establish the relation to prior art.
We can relate operators $a$ and $a^\dagger$ to a pair of ordinary
Bose operators $b$ and $b^\dagger$as follows. Let us define
\begin{equation}\label{hw-osc}
\begin{split}
& N = b^\dagger b ~,~~~~~~~~~~~~ \left[ b,b^\dagger\right]=  1 \\
& \left[ N,b\right] =   -b~,~~~~~~
\left[ N,b^\dagger\right] = b^\dagger~. \\
\end{split}
\end{equation}
The function $\phi(N)$ can be written in a factorized form
\begin{equation}
\phi(N) = \frac{N+\nu}{N+\nu -1} N (N+2\nu -1)~.
\end{equation}
The mapping to Bose operators is given as  follows, \cite{silvio},
\begin{equation}\label{eq:deformation-of-alg}
\begin{split}
a^\dagger = & \sqrt{\frac{\phi(N)}{N}}b^\dagger = b^\dagger \sqrt{\frac{\phi(N+1)}{N+1}} \\
a  = & b \sqrt{\frac{\phi(N)}{N}}  =  \sqrt{\frac{\phi(N+1)}{N+1}} b~. \\
\end{split}
\end{equation}
Then the Hamiltonian takes the form
\begin{equation}
H_\mathrm{SPT} = \frac{\alpha^2}{2} \left(b^\dagger b +\nu \right)^2~.
\end{equation}
This is precisely the results found in Ref.\ \cite{dong+lemus} for the P\"oschl-Teller model.
For the oscillator in the minimal uncertainty length quantum mechanics we must include
the additive constant $1/2\beta$
\begin{equation}
\begin{split}
H_\mathrm{OSC} = & \frac{\beta\omega^2}{2} \left( b^\dagger b +\nu\right)^2 -\frac{1}{2\beta} \\
= & \frac{\beta}{2} N^2 + \frac{\beta\nu}{2}\left( N+\frac{1}{2}\right)~. \\
\end{split}
\end{equation}
This is precisely the result obtained in Ref.\ \cite{meljanac}.

We can also view our result as a deformation of $SU(1,1)$ algebra. Let operators
$K_\pm$ and $K_0$ be the generators of an ordinary undeformed $SU(1,1)$ algebra.
They satisfy commutator relations
$[K_0,K_\pm] = \pm K_\pm$ and $[K_-,K_+] = 2K_0$.
It is well known that the $SU(1,1)$ algebra can be constructed
by a suitable deformation of the ordinary single boson oscillator algebra.
For example,
\begin{equation}
\begin{split}
K_+ = & \sqrt{N+2\nu -1}~ b^\dagger = b^\dagger \sqrt{N+2\nu}~,\\
K_- = & b \sqrt{N+2\nu -1} = \sqrt{N+2\nu}~b~,\\
K_0 = & N + \nu~.\\
\end{split}
\end{equation}
We note that the algebra of $K$ operators is characterized uniquely by the parameter $\nu$ that determines the strength of the P\"oschl-Teller potential, because the Casimir operator is given by
$C_K = \nu(\nu-1)$.
We can write Eq.\ (\ref{eq:deformation-of-alg}) in terms of the $SU(1,1)$ generators
\begin{equation}
\begin{split}
a^\dagger = & \sqrt{\frac{N+\nu}{N+\nu-1}} K_+ =  K_+ \sqrt{\frac{N+\nu+1}{N+\nu}} \\
a^\dagger = &  K_- \sqrt{\frac{N+\nu}{N+\nu-1}} =   \sqrt{\frac{N+\nu+1}{N+\nu}} K_- \\
H = & \frac{\alpha^2}{2}K_0^2~.
\end{split}
\end{equation}
This the result obtained in \cite{dong+lemus}.
We must stress again that $SU(1,1)$ appearing here is not a dynamical symmetry group!

The dynamical symmetry group is simply the dynamical $SU(1,1)$ of the Bose oscillator $b$ and $b^\dagger$.
We have
$S_+ = \frac{1}{2}{b^\dagger}^2$, $S_- = \frac{1}{2} b^2$ and $S_0 = \frac{1}{4} (b b^\dagger + b^\dagger b)$.
States are divided into two infinite dimensional representations of this $SU(1,1)$, as explained earlier.
Even $n$ states belong to the Bargman index $k=1/4$ representation and odd $n$ states belong to the $k=3/4$ representation.
Deformation does not mix these two representations.
We can also define a deformed algebra quadratic in operators $a$ and $a^\dagger$ as follows. Let
$T_- = \frac{1}{2} a^2$, $T_+ = \frac{1}{2}{a^\dagger}^2$ and $T_0 = S_0$.
Operators $T_\pm$ and $T_0$ satisfy commutators of the form
$[T_0,T_\pm] = \pm T_\pm$ and $[T_-,T_+] = G(T_0)$. It is not hard to work out the explicit form of
$T_\pm$ and $G(T_0)$. However, the corresponding quadratic Casimir operator is zero and offers no new information.

In a way this completes the story of how to quantize the system in the deformed quantum mechanics.
Simply search for an undeformed system that is acceptable both to the
deformed commutator $[a,a^\dagger] = 1+2H_0$ and to the Hamiltonian $H$ of the model studied.
This task however may not be easy to carry out. The moral of the story is
that the deformed theory should probably considered as a constrained system and the
quantization must then be carried according to the rules for quantization of constrained systems
explained by Dirac \cite{dirac}.

\section{\label{sec7} The $D$-Dimensional Isotropic Harmonic Oscillator}

In this section we consider the $D$-dimensional extension of the minimal length uncertainty quantum mechanics.
The problem is quite interesting because the extension to higher dimensions implies that spatial
coordinates do not commute \cite{kempf2}. It is a lucky circumstance that rotational symmetry is preserved.
This means that isotropic systems can be reduced to a quantization of some effective
one-dimensional model on the positive real line.

In $D$-spatial dimensions the deformed basic commutators are given by
\begin{equation}\label{3d-deformation}
\begin{split}
\left[ x_i,p_j\right] = & i\left(1+\beta p^2\right)\delta_{ij} + i\beta^\prime p_ip_j \\
\left[ p_i,p_j\right] = & 0\\
\left[ x_i,x_j\right] = & - i\left( (2\beta - \beta^\prime)+(2\beta+\beta^\prime)p^2 \right)L_{ij}~,
\end{split}
\end{equation}
where $L_{ij} = \frac{1}{1+\beta p^2}(x_i p_j -x_j p_i)$ are the components of the angular momentum tensor   \cite{kempf+mm,minic-atall} .
The third commutator in Eq. (\ref{3d-deformation}) implies that the spatial coordinates are noncommutative.
Here we follow the notation and conventions of \cite{minic-atall} where the
the Schr\"odinger equation for the oscillator model was solved.

The momentum space representation is available. In the momentum space the momentum operators can
be represented as simple multiplication.
The rotational symmetry implies that the radial
variable $p = \sqrt{\displaystyle{\sum_{i=1}^D}p_i^2}$ is a good variable.
The position operator is represented by
\begin{equation}
x_i =  i(1+\beta p^2)\frac{\partial}{\partial p_i}
+ i\beta^\prime p_i p_j \frac{\partial}{\partial p_j}
+ i\gamma p_i~.
\end{equation}
Variables $p_i$, $i=1,2,\ldots, D$, run from $-\infty$ to $\infty$.
The measure is given by
\begin{equation}
d\mu = V_{D-1} (1+(\beta+\beta^\prime)p^2)^{\alpha -1}p^{D-1} dp~,
\end{equation}
where $V_{D-1}$ is a volume of the $D-1$ dimensional sphere and  $0\leq p \leq\infty$. The constant  $\alpha$ in the measure is given by
$\alpha = \frac{\gamma}{\beta+\beta^\prime} -\frac{\beta^\prime}{\beta+\beta^\prime}\frac{D-1}{2}$.
Because of the rotational symmetry the square of the operator ${\displaystyle\sum_{i=1}^D}x_i^2$
will involve the $D$-dimensional Laplace operator that can be expressed in spherically symmetric coordinates
$\nabla^2_p = \frac{\partial^2}{\partial p^2} + \frac{D-1}{p}\frac{\partial}{\partial p}
-\frac{l(l+D-2)}{p^2}$,
where angular momentum quantum number is an integer, $l=0,1,\ldots,$.
There is a usual degeneracy in the magnetic quantum number.
The wave functions is factorized into a radial and angular part according to
$\Psi(p_i) =\Psi(p)Y(\Omega)$ where $Y(\Omega)$ is a $D$-dimensional generalization of
spherical harmonics.
In writing the decomposition of the wave function we have suppressed the angular quantum number $l$.
The Hamiltonian of the isotropic $D$-dimensional oscillator is given by
\begin{equation}
H =  \sum_{i=1}^D\left( \frac{1}{2m}p^2_i +\frac{m\omega^2}{2} x^2_i\right) =
\frac{1}{2m} p^2 + \frac{m\omega^2}{2} x^2~,
\end{equation}
where the expression after the second equality sign is given in radial coordinates.
The operator $x^2$ reads explicitly (we use the shorthand notation $L^2 = l(l+D-2)$),
{\small
\begin{equation}
\begin{split}
 - & x^2 =
 \left((1+(\beta + \beta^\prime)p^2)\frac{d}{dp} \right)^2 - \frac{L^2}{p^2} + (\gamma D - 2\beta L^2)\\
 + &
 \left(\frac{D-1}{p} + ((D-1)\beta + 2\gamma)p \right)
 \left(1+(\beta + \beta^\prime)p^2 \right)\frac{d}{dp} \\
 +&  \left(\gamma(\beta D +\beta^\prime + \gamma) - \beta^2 L^2\right)p^2~.
\end{split}
\end{equation}
}

As in the one-dimensional case we work with the Green's function $G(z)$.
The transformation from the momentum space to wave-vector is given by
\begin{equation}
p = \frac{1}{\sqrt{\beta + \beta^\prime}}\tan \sqrt{\beta +\beta^\prime}\rho,
\end{equation}
see Ref.\ \cite{minic-atall}.
The wave-vector now runs over positive values, $0\leq \rho\leq \frac{\pi}{2\sqrt{\beta+\beta^\prime}}$.
The similarity transformation $\Psi(\rho) = J\tilde{\psi}(\rho)$ with
$J = (\cos \sqrt{\beta+\beta^\prime}\rho)^{\gamma/(\beta + \beta^\prime)}$ removes all dependence on the
parameter $\gamma$.
The Green's operator matrix element reads
{\small
\begin{equation}
\begin{split}
& G^{-1}(z)  = \\
& \int\limits^{\frac{\pi}{2\sqrt{\beta+\beta^\prime}}}_0 d\mu(\rho)~
\tilde{\phi}^\ast(\rho)
\left[ z-
\frac{\left(\tan\sqrt{\beta+\beta^\prime}\rho \right)^2}{2m (\beta + \beta^\prime)}   -
\frac{m\omega^2}{2} {x^\prime}^2
\right]
\tilde{\psi}(\rho)~, \\
\end{split}
\end{equation}
}
where the measure is
\begin{equation}
d\mu(\rho) =  \left(\tan\sqrt{\beta+\beta^\prime}\rho \right)^{D-1}
\frac{(\beta +\beta^\prime)^{(1-D)/2} d\rho}{\left(\cos\sqrt{\beta+\beta^\prime}\rho \right)^{2\delta}}
\end{equation}
and $\delta = - \frac{\beta^\prime}{\beta+\beta^\prime} \frac{D-1}{2}$.
The operator ${x^\prime}^2$ is given by
{\small
\begin{equation}
\begin{split}
- & {x^\prime}^2 = \frac{d^2}{d\rho^2}  -2\beta L^2 \\
- & \frac{L^2 (\beta+\beta^\prime)}{\tan^2\sqrt{\beta+\beta^\prime}\rho}
- \frac{\beta^2L^2}{\beta+\beta^\prime}\tan^2\sqrt{\beta+\beta^\prime}\rho \\
+ & \frac{(D-1)\sqrt{\beta+\beta^\prime}}{\tan\sqrt{\beta+\beta^\prime}\rho}
\left(1 + \frac{\beta}{\beta+\beta^\prime}\tan^2\sqrt{\beta+\beta^\prime}\rho\right) \frac{d}{d\rho}~.
\end{split}
\end{equation}
}
Operator ${x^\prime}^2$ involves the term with the first derivative.
This is typical for higher dimensional theories. Such term can be eliminated
by another similarity transformation, $\tilde{\psi}(\rho) = J\psi(\rho)$,
where
$J = \left(
\frac{(\cos\sqrt{\beta+\beta^\prime}\rho)^{\beta/(\beta+\beta^\prime)}}{\sin\sqrt{\beta+\beta^\prime}\rho}
\right)^{(D-1)/2}$.
Finally, we arrive at the Green's function written in terms of
the Hamiltonian of the equivalent P\"oschl-Teller model
\begin{equation}
G^{-1}(z) = \mathrm{constant}\times\int\limits_0^{\pi/2\alpha} dx~
\phi^\ast(x) \left( z^\prime - H_\mathrm{PT}\right) \psi(x)~,
\end{equation}
where
\begin{equation}\label{3d-pt}
H_\mathrm{PT} = \frac{1}{2} p^2 + \frac{(2\alpha)^2}{8} \frac{\nu(\nu -1)}{\cos^2\alpha x} +
\frac{(2\alpha)^2}{8} \frac{\mu(\mu -1)}{\sin^2\alpha x}~.
\end{equation}
In writing the Hamiltonian in equation (\ref{3d-pt})
we have rescaled the variable $\rho$ and have defined several new constants, and have also
performed a shift in the energy variable $z$.
The following definitions apply:
\begin{equation}\label{eq:stdpt-const}
\begin{split}
 \sqrt{\beta+\beta^\prime} \rho = & \alpha x~,~~~~~~
(\beta + \beta^\prime )m\omega^2 = \alpha^2\\
j = & l + \frac{D-3}{2}~,~~~l = 0,1,\cdots,~~~~~~ p = i\frac{d}{d\rho} \\
\nu (\nu-1) = & \frac{\beta^2}{(\beta+\beta^\prime)^2} j(j+1) -
\frac{\beta\beta^\prime}{(\beta+\beta^\prime)^2} \frac{D-1}{2} \\
+ & \frac{1}{m^2\omega^2(\beta+\beta^\prime)^2}
\\
\mu(\mu-1) = & j(j+1) \\
z^\prime = & z +  \frac{1}{2m(\beta+\beta^\prime)} \\
- & \frac{m\omega^2\beta^\prime(\beta^\prime + 2\beta)}{2(\beta+\beta^\prime)}
\left(j(j+1) + \frac{D-1}{2} \right)~.\\
\end{split}
\end{equation}
We should note the following about Eq.\ (\ref{eq:stdpt-const}).
In the definition of the parameter $\nu$, the first two terms originate from the oscillator potential energy. The third term originates from the oscillator kinetic energy. The parameter $\mu$ also originates completely from the oscillator potential energy term.
In the limit, $D=1$, $l=0$ and $\beta^\prime=0$,
the result reduces to the one-dimensional case; in the same limit the parameter $\mu$ becomes zero and the energy parameter $z^\prime$ become the same as in the one-dimension.

The natural coordinates for the P\"oschl-Teller model are
\begin{equation}
\begin{split}
 y = & \cos2\alpha x \\
 k = & -\alpha \{ p, \sin2\alpha x\}~.\\
\end{split}
\end{equation}
After some calculation we arrive at the Heisenberg-Weyl algebra in natural coordinates
\begin{equation}\label{eq:pt-algebra-natural}
\begin{split}
\left[ y,k \right] = & i(2\alpha)^2(1-y^2) \\
\left[ H,y \right] = & -ik \\
\left[ H,k \right] = & i(2\alpha)^2\left(2yH - ik +\frac{(2\alpha)^2}{4}y \right) \\
& + \frac{(2\alpha)^2}{4}\left(\nu(\nu -1) - \mu(\mu -1) \right)~.\\
\end{split}
\end{equation}
This algebra is closed and similar to that of the symmetric P\"oschl-Teller model of
Section \ref{sec4}. However, there is an important difference, the central term in the $[H,k]$
commutator. This means that ladder operators will contain an extra term in addition to a linear combination
of $y$ and $k$ operators.
Following the construction outlined in Section \ref{sec4} we obtain
\begin{equation}
\begin{split}
 a= & \frac{1}{4\alpha^2}
 \left[ y (2\alpha \sqrt{2H} + 2\alpha^2)  +ik - \frac{4\alpha^4 C }{2\alpha\sqrt{2H} - 2\alpha^2}
\right]\\
 a^\dagger = & \frac{1}{4\alpha^2}
 \left[  (2\alpha \sqrt{2H} + 2\alpha^2) y  - ik - \frac{4\alpha^4 C }{2\alpha\sqrt{2H} - 2\alpha^2}
\right],\\
\end{split}
\end{equation}
where
\begin{equation}
C = \nu(\nu-1) - \mu(\mu - 1) =
\left( \nu - \frac{1}{2}\right)^2 -
\left( \mu - \frac{1}{2}\right)^2~.
\end{equation}
Ladder operators satisfy the commutator relations
\begin{equation}\label{eq:81}
\begin{split}
[ H, a ] =&  - a \left( 2\alpha\sqrt{2H} - 2\alpha\right) =
- \left( 2\alpha\sqrt{2H} + 2\alpha\right) a \\
[ H, a^\dagger ] =&   \left( 2\alpha\sqrt{2H} - 2\alpha\right) a^\dagger =
 a^\dagger \left( 2\alpha\sqrt{2H} + 2\alpha\right)~.  \\
\end{split}
\end{equation}
The second set of equalities
follows from commutator relations also satisfied by ladder operators
\begin{equation}\label{eq:84}
\begin{split}
[ \sqrt{2H} , a ] = & -2\alpha a \\
[ \sqrt{2H} , a^\dagger ] = & 2\alpha a^\dagger~, \\
\end{split}
\end{equation}
which indicates that the square root of the Hamiltonian behaves as a natural number operator
for $a$ and $a^\dagger$.

The central term makes appearance in operator products $a^\dagger a$ and $a a^\dagger$, or in the
commutator $[a,a^\dagger]$ and the anti-commutator $\{ a,a^\dagger\}$.
After tedious calculations we find
\begin{equation}
\begin{split}
a^\dagger a = \phi(\sqrt{2H}) = & - \frac{Q}{2}\frac{\frac{\sqrt{2H}}{2\alpha}}{\frac{\sqrt{2H}}{2\alpha} -1} +
\left( \frac{\sqrt{2H}}{2\alpha}\right)^2 \\
+ &\frac{C^2}{4} \frac{\frac{\sqrt{2H}}{2\alpha}}
{\left( \frac{\sqrt{2H}}{2\alpha}-1\right)\left(2 \frac{\sqrt{2H}}{2\alpha} -1 \right)^2}\\
a a^\dagger = & \phi(\sqrt{2H} +2\alpha)~,\\
\end{split}
\end{equation}
where
\begin{equation}
Q = \nu(\nu-1)+\mu(\mu -1) = \left(\nu-\frac{1}{2}\right)^2 +
\left(\mu - \frac{1}{2}\right)^2 - \frac{1}{2}~.
\end{equation}
The ground state of the model is defined by
\begin{equation}\label{eq:ptgs}
a \vert 0;\nu,\mu\rangle = 0~.
\end{equation}

The existence of the ground state solution to equation (\ref{eq:ptgs}) implies some restrictions on
identification of the square root of the Hamiltonian with the number operator.
Using Eq.\ (\ref{eq:84}) we define
\begin{equation}
\frac{\sqrt{2H}}{2\alpha} = N + d~,
\end{equation}
where $N$ is an ordinary number operator for operators $a$ and $a^\dagger$ defined by
$[N,a] = -a$ and $[N,a^\dagger] = a^\dagger$ and $d$ is some constant. The existence of the
ground state solution determines the possible values of parameter $d$. We have
{\small
\begin{equation}
\phi(N=0)=0 = - \frac{Q}{2}\frac{d}{d -1} + d^2  + \frac{C^2}{4} \frac{d}{(d-1)(2d-1)^2}~.
\end{equation}
}
There are five solutions, $d=0, \frac{\nu+\mu}{2}, \frac{\nu + 1-\mu}{2},
\frac{1-\nu + \mu}{2}$ and $\frac{2-\nu-\mu}{2}$. With $d=0$, the first excited state is infinite. The third, the fourth and the fifth solutions appear to correspond to different
parameterizations of the potential of the model. We work with $d= \frac{\nu+\mu}{2}$. In general,
if the combination $(\nu+\mu)/2$ is not a positive integer, then there will be an infinite
tower of states. The function $\phi(\sqrt{2H})$ factorizes
($\sqrt{2H} = 2\alpha \left(N + \frac{\nu+\mu}{2} \right)$),
\begin{equation}
\begin{split}
\phi(N) = & N (N+\nu +\mu -1) \times \\
& \frac{(N+\frac{2\nu -1}{2})(N+\frac{2\mu -1}{2})(N+\frac{\nu+\mu}{2})}
{(N+\frac{\nu+\mu-1}{2})^2(N+\frac{\nu+\mu-2}{2})} ~.\\
\end{split}
\end{equation}
The energy spectrum is given by
\begin{equation}
\begin{split}
E_n = & 2\alpha^2 \left( n + \frac{\nu + \mu}{2} \right)^2~,~~~~~~ n = 0,1,2,\cdots, \\
\vert n;&\nu,\mu\rangle \propto         \left(a^\dagger\right)^n\vert 0;\nu,\mu\rangle~,
\end{split}
\end{equation}
where we assumed that the ground state is normalized to unity.

One can show, using the explicit expression for ladder operators in terms of natural operators
$y$ and $k$, that energy eigenfunctions are in fact Jacobi polynomials $P_{(n-l)/2}^{(\nu-1/2, \mu-1/2)}(y)$ \cite{gradshteyn}.
We also reproduce the energy formula given in Ref.\ \cite{minic-atall} for the $D$-dimensional isotropic oscillator with minimal length uncertainty quantum mechanics
{\small
\begin{equation}
\begin{split}
& E_{n,l} =  \omega
\left( n+ \frac{D}{2}\right) \times \\
& \sqrt{1+m^2\omega^2\left[\beta^2 j(j+1) +
\frac{\beta^2 + {\beta^\prime}^2 - 2\beta\beta^\prime(D-3)}{4}\right]} \\
+ & \frac{m\omega^2 (\beta+\beta^\prime)}{2} \left( n+ \frac{D}{2}\right)^2 +
\frac{m\omega^2 \beta^\prime D}{4}
\\
+ &
\frac{m\omega^2(\beta-\beta^\prime)}{2} \left(j(j+1) - \frac{(D-1)(D-3)}{4} \right)~.
\end{split}
\end{equation}
}

Just like in one-dimension, the model can be described by a deformation of a single constrained
boson of ordinary one-dimensional quantum mechanics. The constraint comes in the form of the boson
coordinate space being restricted to only a segment of the real line.  It is also interesting that the
$D$-dimensional model appears to allow finite dimensional representations, see [16].
These deserve to be explored in more detail.

\section{\label{sec8}Summary and outlook}

In this paper we have studied harmonic oscillator models in the quantum theory
with minimal length uncertainty relations.
Such models may be relevant to quantum gravity at the Planck scale or
may appear as an effective theory where modified uncertainty relations
are introduced to capture certain features of physics below some scale.
Examples of the second kind are models of rotational and vibrational spectra of molecules in
molecular and chemical physics and in heavy deformed nuclei in nuclear physics.

We have focused on using of operator techniques because modification of the
basic commutator of the quantum theory implies  appearance of minimal
length and more generally minimal momentum. This then means
that the position and the momentum operators cannot be diagonalized any more.
Consequently, the Schr\"odinger equation as a differential or integral equation becomes unavailable.
We have shown that that operator techniques work and that complete knowledge of
the system can be obtained. The next obvious step would be make generalizations and to consider models more
complicated than the oscillator. The Coulomb problem is the first important candidate.
Also, application to the field theory is desired too.

It is also of interest to
learn how operator techniques extensions work when both, the position and momentum are
limited by minimal uncertainties. This problem is in part related to $q$-oscillators.
There exists a great deal of literature on $q$-deformed oscillators.
However, the present problem is more general than the typical $q$-oscillators.
First, the symmetry principles place constraints on the form  that the basic commutator can take and
this selects the applicable $q$-oscillators.
For example, in Ref.\ \cite{smolin}, an extension of the special relativity that incorporates
the minimal invariant length and the minimal invariant momentum was constructed and
in this particular extension of the basic commutator of quantum mechanics is deformed
roughly as,  $\alpha x^2 + \beta p^2 + \gamma (xp+px)$.
The second part involves of the present problem then involves the diagonalization of an
arbitrary system in the basis defined by a $q$-oscillator. This is a difficult problem.

Another interesting fact that follows from the present work is that in both cases we can understand the
quantization in the deformed theory as a deformed  ordinary Bose oscillator.
Perhaps this means that the deformed quantum mechanics
is in fact an ordinary quantum mechanics with very complicated constraints. In that case the
Dirac's theory of quantization with constraints, \cite{dirac} may be an answer.

In closing let us also mention that the problem of the oscillator in a constant external field can also
be incorporated in the formalism. The constant external field simply adds a term of the form
$H_{int} = gx$ to the Hamiltonian. Once the models are transformed to P\"oschl-Teller form there will be an extra
term with the first derivative present. Such term can always be removed by an appropriate
similarity transformation. Once this transformation is carried out, the analysis goes through as
described in this paper.

\begin{acknowledgments}
This work has been supported in part by the Research Corporation.
\end{acknowledgments}


\end{document}